# Displaying perfusion MRI images as color intensity projections

**F. Hoefnagels[1], K. S. Cover[2], E. Sanchez[3], and F. J. Lagerwaard[4]**
[1]Neurosurgery, [2]Physics and Medical Technology, and [3]Radiology and
[4]Radiation Oncology, VU University Medical Center, Amsterdam, Netherlands

## Abstract

Dynamic susceptibility-weighted contrast-enhanced (DSC) MRI or perfusion-MRI plays an important role in the non-invasive assessment of tumor vascularity. However, the large number of images provided by the method makes display and interpretation of the results challenging. Current practice is to display the perfusion information as relative cerebral blood volume maps (rCBV). Color intensity projections (CIPs) provides a simple, intuitive display of the perfusion-MRI data so that regional perfusion characteristics are intrinsically integrated into the anatomy structure the T2 images. The ease of use and quick calculation time of CIPs should allow it to be easily integrated into current analysis and interpretation pipelines.

## Introduction

Dynamic susceptibility-weighted contrast-enhanced (DSC) (also called perfusion MRI) is capable of quantifying microvessel density (vascularity) and permeability of brain tissue by assessment of the relative cerebral blood volume map (rCBV) [1-2]. Perfusion MRI has been reported to be of diagnostic and prognostic value in both glioma and brain metastases patients. In addition, perfusion MRI can be used for differentiating radiation-induced necrosis from tumor progression. However, the difficulties in 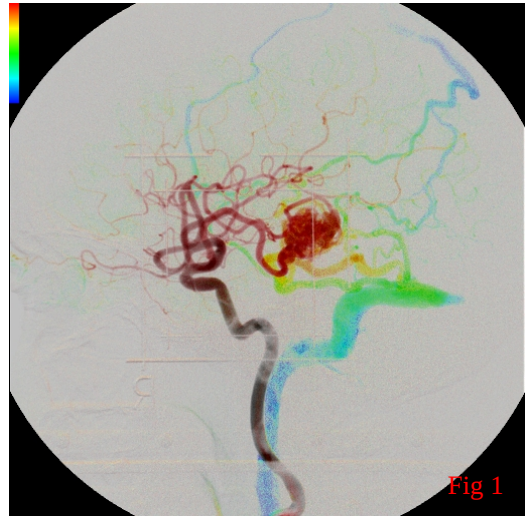 displaying and interpreting perfusion MRI images have inhibited its widespread use. The method of color intensity projections (CIPs) is proposed for displaying a series of perfusion MRI images, including the CBV information, into a single color image. CIPs have been used previously to display motion in 4DCTs [3] and flow in digital subtraction angiography [4] (Fig 1).

## Method

A 52 year old female underwent radiosurgery for a single cerebral metastasis in the left frontal lobe. On follow-up imaging 6 months post-radiosurgery there was increased gadolinium enhancement at the lateral border of the treated lesion (Figure 2g), with a differential diagnosis of radiation-induced necrosis or tumor progression. A perfusion MRI was performed with a T2 echo-planar sequence after the injection of gadolinium. The images from immediately before the appearance of contrast in the brain until 23 seconds later were included in the time series used to generate the CIP. The CIP image was calculated on a pixel by pixel basis and took less than 2 seconds to calculate. In a CIP image, a grayscale pixel indicates a constant pixel value over the time series while a colored pixel indicates a varying pixel value. Thus, pixels appearing as grayscale indicate no perfusion. The larger the range of the pixel values the stronger the color, also known as color saturation. The hue of the color (red-yellow-green-blue) indicates whether the average value of the pixel was closer to the minimum (red) or maximum (blue) pixel intensity over the time series. As the CBV is directly proportional to the average pixel value, the hue of the CIPs encodes the CBV. Thus, vessels appear in red because contrast quickly fills the vessels yielding a low average pixel value. The green and yellow colors indicate highly perfused tissue while blue pixels indicate areas with low perfusion. For comparison, the rCBV map was calculated with the Leonardo VD10B Syno software.

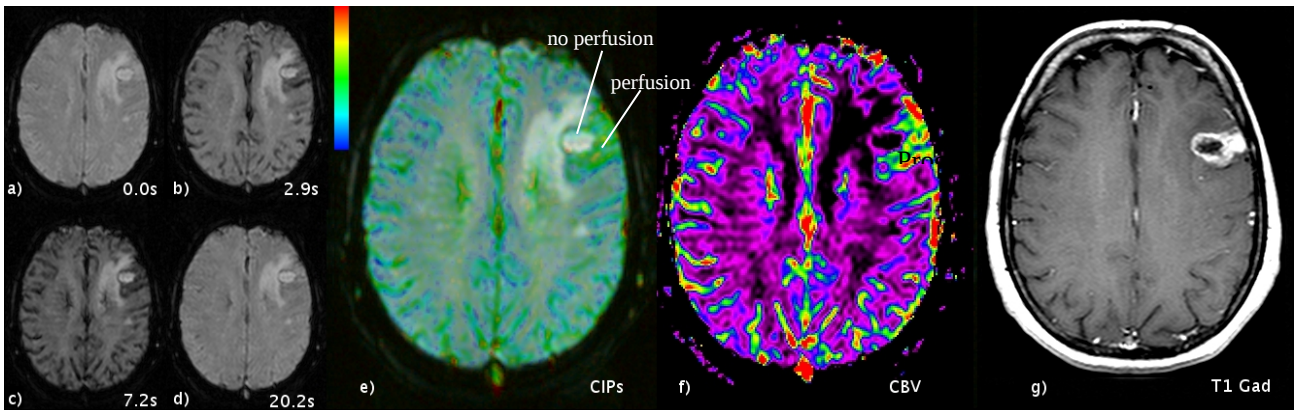

*Figure 2. Brain metastasis treated with radiosurgery (a-d) perfusion MRI, (e) color intensity projection (CIPs), (f) rCBV map, and (g) gadolinium enhanced T1 weighted MRI.*

**Results**

Both the CIP image and rCBV map reveal relatively high perfusion on the lateral and dorsal aspect of the lesion, corresponding to the contrast-enhancement on the gadolinium-enhanced T1 images, indicative of tumor progression. The advantage of the CIP image over the rCBV map is that the former shows the color coding integrated with the T2 MRI anatomical structures, which can make the interpretation of regional differences in perfusion more straightforward.

**Discussion**

CIP of perfusion MRI images provides a fast and practical summary of relative regional brain perfusion. The major advantage CIPs has over rCBV maps is the intrinsic integration of the T2 MRI anatomy into the image, enabling rapid pathology and anatomical structures, such as vessels. A qualitative comparative study between the rCBV maps and CIPs in patients with radiosurgically treated brain metastases is currently ongoing.

**Acknowledgements**

This work was funded by the Department of Radiotherapy, the Department of Physics and Medical Technology and the Department of Neurosurgery at the VU University Medical Center, Amsterdam, The Netherlands. Additional funding was received from the Netherlands' Virtual